\documentclass[journal]{IEEEtran}
\usepackage{hyperref}
\usepackage{url}

\usepackage{times}
\usepackage[pdftex]{graphicx} 
\usepackage{subfigure} 
\usepackage{amsmath,epsfig,cite,mathtools} 
\usepackage{amsfonts}
\usepackage{amssymb}
\usepackage{amsthm}
\usepackage{mathrsfs}
\usepackage[ruled,vlined]{algorithm2e}
\usepackage{amsmath}
\usepackage{paralist}
\usepackage{booktabs}
\usepackage{color}
\usepackage{url}
\usepackage{graphicx}
\DeclareGraphicsExtensions{.pdf,.jpeg,.png,.epbibdesk}
\usepackage{epstopdf}
\usepackage[T1]{fontenc}
\usepackage{graphicx}
\usepackage{yhmath}
\usepackage{hyperref}
\usepackage{amsthm}
\usepackage{enumitem}
\usepackage{balance}

\usepackage{amssymb}
\usepackage{amsfonts}
\usepackage{mathrsfs}
\usepackage{xspace}
\usepackage{bm}
\usepackage{upgreek}

\newcommand{\safemath}[2]{\newcommand{#1}{\ensuremath{#2}\xspace}}

\safemath{\bma}{\mathbf{a}}
\safemath{\bmb}{\mathbf{b}}
\safemath{\bmc}{\mathbf{c}}
\safemath{\bmd}{\mathbf{d}}
\safemath{\bme}{\mathbf{e}}
\safemath{\bmf}{\mathbf{f}}
\safemath{\bmg}{\mathbf{g}}
\safemath{\bmh}{\mathbf{h}}
\safemath{\bmi}{\mathbf{i}}
\safemath{\bmj}{\mathbf{j}}
\safemath{\bmk}{\mathbf{k}}
\safemath{\bml}{\mathbf{l}}
\safemath{\bmm}{\mathbf{m}}
\safemath{\bmn}{\mathbf{n}}
\safemath{\bmo}{\mathbf{o}}
\safemath{\bmp}{\mathbf{p}}
\safemath{\bmq}{\mathbf{q}}
\safemath{\bmr}{\mathbf{r}}
\safemath{\bms}{\mathbf{s}}
\safemath{\bmt}{\mathbf{t}}
\safemath{\bmu}{\mathbf{u}}
\safemath{\bmv}{\mathbf{v}}
\safemath{\bmw}{\mathbf{w}}
\safemath{\bmx}{\mathbf{x}}
\safemath{\bmy}{\mathbf{y}}
\safemath{\bmz}{\mathbf{z}}
\safemath{\bmzero}{\mathbf{0}}
\safemath{\bmone}{\mathbf{1}}

\bmdefine{\biad}{a}
\bmdefine{\bibd}{b}
\bmdefine{\bicd}{c}
\bmdefine{\bidd}{d}
\bmdefine{\bied}{e}
\bmdefine{\bifd}{f}
\bmdefine{\bigd}{g}
\bmdefine{\bihd}{h}
\bmdefine{\biid}{i}
\bmdefine{\bijd}{j}
\bmdefine{\bikd}{k}
\bmdefine{\bild}{l}
\bmdefine{\bimd}{m}
\bmdefine{\bind}{n}
\bmdefine{\biod}{o}
\bmdefine{\bipd}{p}
\bmdefine{\biqd}{q}
\bmdefine{\bird}{r}
\bmdefine{\bisd}{s}
\bmdefine{\bitd}{t}
\bmdefine{\biud}{u}
\bmdefine{\bivd}{v}
\bmdefine{\biwd}{w}
\bmdefine{\bixd}{x}
\bmdefine{\biyd}{y}
\bmdefine{\bizd}{z}

\bmdefine{\bixid}{\xi}
\bmdefine{\bilambdad}{\lambda}
\bmdefine{\bimud}{\mu}
\bmdefine{\bithetad}{\theta}
\bmdefine{\biphid}{\phi}
\bmdefine{\bideltad}{\delta}

\safemath{\bmia}{\biad}
\safemath{\bmib}{\bibd}
\safemath{\bmic}{\bicd}
\safemath{\bmid}{\bidd}
\safemath{\bmie}{\bied}
\safemath{\bmif}{\bifd}
\safemath{\bmig}{\bigd}
\safemath{\bmih}{\bihd}
\safemath{\bmii}{\biid}
\safemath{\bmij}{\bijd}
\safemath{\bmik}{\bikd}
\safemath{\bmil}{\bild}
\safemath{\bmim}{\bimd}
\safemath{\bmin}{\bind}
\safemath{\bmio}{\biod}
\safemath{\bmip}{\bipd}
\safemath{\bmiq}{\biqd}
\safemath{\bmir}{\bird}
\safemath{\bmis}{\bisd}
\safemath{\bmit}{\bitd}
\safemath{\bmiu}{\biud}
\safemath{\bmiv}{\bivd}
\safemath{\bmiw}{\biwd}
\safemath{\bmix}{\bixd}
\safemath{\bmiy}{\biyd}
\safemath{\bmiz}{\bizd}

\safemath{\bmxi}{\bixid}
\safemath{\bmlambda}{\bilambdad}
\safemath{\bmmu}{\bimud}
\safemath{\bmtheta}{\bithetad}
\safemath{\bmphi}{\biphid}
\safemath{\bmdelta}{\bideltad}

\safemath{\bA}{\mathbf{A}}
\safemath{\bB}{\mathbf{B}}
\safemath{\bC}{\mathbf{C}}
\safemath{\bD}{\mathbf{D}}
\safemath{\bE}{\mathbf{E}}
\safemath{\bF}{\mathbf{F}}
\safemath{\bG}{\mathbf{G}}
\safemath{\bH}{\mathbf{H}}
\safemath{\bI}{\mathbf{I}}
\safemath{\bJ}{\mathbf{J}}
\safemath{\bK}{\mathbf{K}}
\safemath{\bL}{\mathbf{L}}
\safemath{\bM}{\mathbf{M}}
\safemath{\bN}{\mathbf{N}}
\safemath{\bO}{\mathbf{O}}
\safemath{\bP}{\mathbf{P}}
\safemath{\bQ}{\mathbf{Q}}
\safemath{\bR}{\mathbf{R}}
\safemath{\bS}{\mathbf{S}}
\safemath{\bT}{\mathbf{T}}
\safemath{\bU}{\mathbf{U}}
\safemath{\bV}{\mathbf{V}}
\safemath{\bW}{\mathbf{W}}
\safemath{\bX}{\mathbf{X}}
\safemath{\bY}{\mathbf{Y}}
\safemath{\bZ}{\mathbf{Z}}

\safemath{\bZero}{\mathbf{0}}
\safemath{\bOne}{\mathbf{1}}
\safemath{\bDelta}{\mathbf{\Delta}}
\safemath{\bLambda}{\mathbf{\UpLambda}}
\safemath{\bPhi}{\mathbf{\Upphi}}
\safemath{\bSigma}{\mathbf{\Upsigma}}
\safemath{\bOmega}{\mathbf{\Upomega}}
\safemath{\bTheta}{\mathbf{\Uptheta}}

\bmdefine{\biAd}{A}
\bmdefine{\biBd}{B}
\bmdefine{\biCd}{C}
\bmdefine{\biDd}{D}
\bmdefine{\biEd}{E}
\bmdefine{\biFd}{F}
\bmdefine{\biGd}{G}
\bmdefine{\biHd}{H}
\bmdefine{\biId}{I}
\bmdefine{\biJd}{J}
\bmdefine{\biKd}{K}
\bmdefine{\biLd}{L}
\bmdefine{\biMd}{M}
\bmdefine{\biNd}{N}
\bmdefine{\biOd}{O}
\bmdefine{\biPd}{P}
\bmdefine{\biQd}{Q}
\bmdefine{\biRd}{R}
\bmdefine{\biSd}{S}
\bmdefine{\biTd}{T}
\bmdefine{\biUd}{U}
\bmdefine{\biVd}{V}
\bmdefine{\biWd}{W}
\bmdefine{\biXd}{X}
\bmdefine{\biYd}{Y}
\bmdefine{\biZd}{Z}

\bmdefine{\biDelta}{\Delta}
\bmdefine{\biLambda}{\Lambda}
\bmdefine{\biPhi}{\Phi}
\bmdefine{\biSigma}{\Sigma}
\bmdefine{\biOmega}{\Omega}
\bmdefine{\biTheta}{\Theta}

\safemath{\bimA}{\biAd}
\safemath{\bimB}{\biBd}
\safemath{\bimC}{\biCd}
\safemath{\bimD}{\biDd}
\safemath{\bimE}{\biEd}
\safemath{\bimF}{\biFd}
\safemath{\bimG}{\biGd}
\safemath{\bimH}{\biHd}
\safemath{\bimI}{\biId}
\safemath{\bimJ}{\biJd}
\safemath{\bimK}{\biKd}
\safemath{\bimL}{\biLd}
\safemath{\bimM}{\biMd}
\safemath{\bimN}{\biNd}
\safemath{\bimO}{\biOd}
\safemath{\bimP}{\biPd}
\safemath{\bimQ}{\biQd}
\safemath{\bimR}{\biRd}
\safemath{\bimS}{\biSd}
\safemath{\bimT}{\biTd}
\safemath{\bimU}{\biUd}
\safemath{\bimV}{\biVd}
\safemath{\bimW}{\biWd}
\safemath{\bimX}{\biXd}
\safemath{\bimY}{\biYd}
\safemath{\bimZ}{\biZd}

\safemath{\bimDelta}{\biDelta}
\safemath{\bimLambda}{\biLambda}
\safemath{\bimPhi}{\biPhi}
\safemath{\bimSigma}{\biSigma}
\safemath{\bimOmega}{\biOmega}
\safemath{\bimTheta}{\biTheta}

\safemath{\setA}{\mathcal{A}}
\safemath{\setB}{\mathcal{B}}
\safemath{\setC}{\mathcal{C}}
\safemath{\setD}{\mathcal{D}}
\safemath{\setE}{\mathcal{E}}
\safemath{\setF}{\mathcal{F}}
\safemath{\setG}{\mathcal{G}}
\safemath{\setH}{\mathcal{H}}
\safemath{\setI}{\mathcal{I}}
\safemath{\setJ}{\mathcal{J}}
\safemath{\setK}{\mathcal{K}}
\safemath{\setL}{\mathcal{L}}
\safemath{\setM}{\mathcal{M}}
\safemath{\setN}{\mathcal{N}}
\safemath{\setO}{\mathcal{O}}
\safemath{\setP}{\mathcal{P}}
\safemath{\setQ}{\mathcal{Q}}
\safemath{\setR}{\mathcal{R}}
\safemath{\setS}{\mathcal{S}}
\safemath{\setT}{\mathcal{T}}
\safemath{\setU}{\mathcal{U}}
\safemath{\setV}{\mathcal{V}}
\safemath{\setW}{\mathcal{W}}
\safemath{\setX}{\mathcal{X}}
\safemath{\setY}{\mathcal{Y}}
\safemath{\setZ}{\mathcal{Z}}
\safemath{\emptySet}{\varnothing}

\safemath{\colA}{\mathscr{A}}
\safemath{\colB}{\mathscr{B}}
\safemath{\colC}{\mathscr{C}}
\safemath{\colD}{\mathscr{D}}
\safemath{\colE}{\mathscr{E}}
\safemath{\colF}{\mathscr{F}}
\safemath{\colG}{\mathscr{G}}
\safemath{\colH}{\mathscr{H}}
\safemath{\colI}{\mathscr{I}}
\safemath{\colJ}{\mathscr{J}}
\safemath{\colK}{\mathscr{K}}
\safemath{\colL}{\mathscr{L}}
\safemath{\colM}{\mathscr{M}}
\safemath{\colN}{\mathscr{N}}
\safemath{\colO}{\mathscr{O}}
\safemath{\colP}{\mathscr{P}}
\safemath{\colQ}{\mathscr{Q}}
\safemath{\colR}{\mathscr{R}}
\safemath{\colS}{\mathscr{S}}
\safemath{\colT}{\mathscr{T}}
\safemath{\colU}{\mathscr{U}}
\safemath{\colV}{\mathscr{V}}
\safemath{\colW}{\mathscr{W}}
\safemath{\colX}{\mathscr{X}}
\safemath{\colY}{\mathscr{Y}}
\safemath{\colZ}{\mathscr{Z}}

\safemath{\opA}{\mathbb{A}}
\safemath{\opB}{\mathbb{B}}
\safemath{\opC}{\mathbb{C}}
\safemath{\opD}{\mathbb{D}}
\safemath{\opE}{\mathbb{E}}
\safemath{\opF}{\mathbb{F}}
\safemath{\opG}{\mathbb{G}}
\safemath{\opH}{\mathbb{H}}
\safemath{\opI}{\mathbb{I}}
\safemath{\opJ}{\mathbb{J}}
\safemath{\opK}{\mathbb{K}}
\safemath{\opL}{\mathbb{L}}
\safemath{\opM}{\mathbb{M}}
\safemath{\opN}{\mathbb{N}}
\safemath{\opO}{\mathbb{O}}
\safemath{\opP}{\mathbb{P}}
\safemath{\opQ}{\mathbb{Q}}
\safemath{\opR}{\mathbb{R}}
\safemath{\opS}{\mathbb{S}}
\safemath{\opT}{\mathbb{T}}
\safemath{\opU}{\mathbb{U}}
\safemath{\opV}{\mathbb{V}}
\safemath{\opW}{\mathbb{W}}
\safemath{\opX}{\mathbb{X}}
\safemath{\opY}{\mathbb{Y}}
\safemath{\opZ}{\mathbb{Z}}
\safemath{\opZero}{\mathbb{O}}
\safemath{\identityop}{\opI}

\safemath{\veca}{\bma}
\safemath{\vecb}{\bmb}
\safemath{\vecc}{\bmc}
\safemath{\vecd}{\bmd}
\safemath{\vece}{\bme}
\safemath{\vecf}{\bmf}
\safemath{\vecg}{\bmg}
\safemath{\vech}{\bmh}
\safemath{\veci}{\bmi}
\safemath{\vecj}{\bmj}
\safemath{\veck}{\bmk}
\safemath{\vecl}{\bml}
\safemath{\vecm}{\bmm}
\safemath{\vecn}{\bmn}
\safemath{\veco}{\bmo}
\safemath{\vecp}{\bmp}
\safemath{\vecq}{\bmq}
\safemath{\vecr}{\bmr}
\safemath{\vecs}{\bms}
\safemath{\vect}{\bmt}
\safemath{\vecu}{\bmu}
\safemath{\vecv}{\bmv}
\safemath{\vecw}{\bmw}
\safemath{\vecx}{\bmx}
\safemath{\vecy}{\bmy}
\safemath{\vecz}{\bmz}

\safemath{\veczero}{\bmzero}
\safemath{\vecone}{\bmone}
\safemath{\vecxi}{\bmxi}
\safemath{\veclambda}{\bmlambda}
\safemath{\vecmu}{\bmmu}
\safemath{\vectheta}{\bmtheta}
\safemath{\vecphi}{\bmphi}
\safemath{\vecdelta}{\bmdelta}

\safemath{\matA}{\bA}
\safemath{\matB}{\bB}
\safemath{\matC}{\bC}
\safemath{\matD}{\bD}
\safemath{\matE}{\bE}
\safemath{\matF}{\bF}
\safemath{\matG}{\bG}
\safemath{\matH}{\bH}
\safemath{\matI}{\bI}
\safemath{\matJ}{\bJ}
\safemath{\matK}{\bK}
\safemath{\matL}{\bL}
\safemath{\matM}{\bM}
\safemath{\matN}{\bN}
\safemath{\matO}{\bO}
\safemath{\matP}{\bP}
\safemath{\matQ}{\bQ}
\safemath{\matR}{\bR}
\safemath{\matS}{\bS}
\safemath{\matT}{\bT}
\safemath{\matU}{\bU}
\safemath{\matV}{\bV}
\safemath{\matW}{\bW}
\safemath{\matX}{\bX}
\safemath{\matY}{\bY}
\safemath{\matZ}{\bZ}
\safemath{\matzero}{\bmzero}

\safemath{\matDelta}{\bDelta}
\safemath{\matLambda}{\bLambda}
\safemath{\matPhi}{\bPhi}
\safemath{\matSigma}{\bSigma}
\safemath{\matOmega}{\bOmega}
\safemath{\matTheta}{\bTheta}

\safemath{\matidentity}{\matI}
\safemath{\matone}{\matO}

\safemath{\rnda}{A}
\safemath{\rndb}{B}
\safemath{\rndc}{C}
\safemath{\rndd}{D}
\safemath{\rnde}{E}
\safemath{\rndf}{F}
\safemath{\rndg}{G}
\safemath{\rndh}{H}
\safemath{\rndi}{I}
\safemath{\rndj}{J}
\safemath{\rndk}{K}
\safemath{\rndl}{L}
\safemath{\rndm}{M}
\safemath{\rndn}{N}
\safemath{\rndo}{O}
\safemath{\rndp}{P}
\safemath{\rndq}{Q}
\safemath{\rndr}{R}
\safemath{\rnds}{S}
\safemath{\rndt}{T}
\safemath{\rndu}{U}
\safemath{\rndv}{V}
\safemath{\rndw}{W}
\safemath{\rndx}{X}
\safemath{\rndy}{Y}
\safemath{\rndz}{Z}

\safemath{\rveca}{\bimA}
\safemath{\rvecb}{\bimB}
\safemath{\rvecc}{\bimC}
\safemath{\rvecd}{\bimD}
\safemath{\rvece}{\bimE}
\safemath{\rvecf}{\bimF}
\safemath{\rvecg}{\bimG}
\safemath{\rvech}{\bimH}
\safemath{\rveci}{\bimI}
\safemath{\rvecj}{\bimJ}
\safemath{\rveck}{\bimK}
\safemath{\rvecl}{\bimL}
\safemath{\rvecm}{\bimM}
\safemath{\rvecn}{\bimN}
\safemath{\rveco}{\bomO}
\safemath{\rvecp}{\bimP}
\safemath{\rvecq}{\bimQ}
\safemath{\rvecr}{\bimR}
\safemath{\rvecs}{\bimS}
\safemath{\rvect}{\bimT}
\safemath{\rvecu}{\bimU}
\safemath{\rvecv}{\bimV}
\safemath{\rvecw}{\bimW}
\safemath{\rvecx}{\bimX}
\safemath{\rvecy}{\bimY}
\safemath{\rvecz}{\bimZ}

\safemath{\rvecxi}{\bmxi}
\safemath{\rveclambda}{\bmlambda}
\safemath{\rvecmu}{\bmmu}
\safemath{\rvectheta}{\bmtheta}
\safemath{\rvecphi}{\bmphi}

\safemath{\rmatA}{\bimA}
\safemath{\rmatB}{\bimB}
\safemath{\rmatC}{\bimC}
\safemath{\rmatD}{\bimD}
\safemath{\rmatE}{\bimE}
\safemath{\rmatF}{\bimF}
\safemath{\rmatG}{\bimG}
\safemath{\rmatH}{\bimH}
\safemath{\rmatI}{\bimI}
\safemath{\rmatJ}{\bimJ}
\safemath{\rmatK}{\bimK}
\safemath{\rmatL}{\bimL}
\safemath{\rmatM}{\bimM}
\safemath{\rmatN}{\bimN}
\safemath{\rmatO}{\bimO}
\safemath{\rmatP}{\bimP}
\safemath{\rmatQ}{\bimQ}
\safemath{\rmatR}{\bimR}
\safemath{\rmatS}{\bimS}
\safemath{\rmatT}{\bimT}
\safemath{\rmatU}{\bimU}
\safemath{\rmatV}{\bimV}
\safemath{\rmatW}{\bimW}
\safemath{\rmatX}{\bimX}
\safemath{\rmatY}{\bimY}
\safemath{\rmatZ}{\bimZ}

\safemath{\rmatDelta}{\bimDelta}
\safemath{\rmatLambda}{\bimLambda}
\safemath{\rmatPhi}{\bimPhi}
\safemath{\rmatSigma}{\bimSigma}
\safemath{\rmatOmega}{\bimOmega}
\safemath{\rmatTheta}{\bimTheta}

\usepackage{amssymb}
\usepackage{amsfonts}
\usepackage{mathrsfs}
\usepackage{xspace}
\usepackage{bm}
\usepackage{fancyref}
\usepackage{textcomp}

\usepackage{multirow}
\usepackage{stmaryrd}

\newenvironment{textbmatrix}{	\setlength{\arraycolsep}{2.5pt}%
								\big[\begin{matrix}}{\end{matrix}\big]%
								\raisebox{0.08ex}{\vphantom{M}}}

\def\be{\begin{equation}}
\def\ee{\end{equation}}
\def\een{\nonumber \end{equation}}
\def\mat{\begin{bmatrix}}
\def\emat{\end{bmatrix}}
\def\btm{\begin{textbmatrix}}
\def\etm{\end{textbmatrix}}

\def\ba#1\ea{\begin{align}#1\end{align}}
\def\bas#1\eas{\begin{align*}#1\end{align*}}
\def\bs#1\es{\begin{split}#1\end{split}} 
\def\bg#1\eg{\begin{gather}#1\end{gather}}
\def\bml#1\eml{\begin{multline}#1\end{multline}}
\def\bi#1\ei{\begin{itemize}#1\end{itemize}}

\safemath{\dirac}{\delta}					
\safemath{\krond}{\dirac}					

\safemath{\upto}{\uparrow}
\safemath{\downto}{\downarrow}
\safemath{\iu}{j}							
\safemath{\ev}{\lambda}						
\safemath{\hilseqspace}{l^{2}}				
\newcommand{\banachfunspace}[1]{\setL^{#1}}	
\safemath{\hilfunspace}{\banachfunspace{2}}	

\safemath{\SNR}{\text{\sc snr}} 				
\safemath{\No}{N_0}							
\safemath{\Es}{E_s}							
\safemath{\Eb}{E_b}							
\safemath{\EbNo}{\frac{\Eb}{\No}}
\safemath{\EsNo}{\frac{\Es}{\No}}

\DeclareMathOperator{\CHop}{\ensuremath{\opH}} 
\safemath{\tvir}{\rndh_{\CHop}}				
\safemath{\tvtf}{\rndl_{\CHop}}				
\safemath{\spf}{\rnds_{\CHop}}				
\safemath{\bff}{H_{\CHop}}					

\safemath{\ircf}{r_{h}}						
\safemath{\tftvcf}{r_{s}}					
\safemath{\tfcf}{r_{l}}						
\safemath{\bfcf}{r_{H}}						

\safemath{\tcorr}{c_h}						
\safemath{\scf}{c_{s}}						
\safemath{\tfcorr}{c_{l}}					
\safemath{\fcorr}{c_{H}}						

\safemath{\mi}{I}							
\safemath{\capacity}{C}						

\safemath{\normal}{\mathcal{N}}			
\safemath{\jpg}{\mathcal{CN}}			
\safemath{\mchain}{\leftrightarrow}		

\safemath{\dB}{\,\mathrm{dB}}
\safemath{\dBm}{\,\mathrm{dBm}}
\safemath{\Hz}{\,\mathrm{Hz}}
\safemath{\kHz}{\,\mathrm{kHz}}
\safemath{\MHz}{\,\mathrm{MHz}}
\safemath{\GHz}{\,\mathrm{GHz}}
\safemath{\s}{\,\mathrm{s}}
\safemath{\ms}{\,\mathrm{ms}}
\safemath{\mus}{\,\mathrm{\text{\textmu}s}}
\safemath{\ns}{\,\mathrm{ns}}
\safemath{\ps}{\,\mathrm{ps}}
\safemath{\meter}{\,\mathrm{m}}
\safemath{\mm}{\,\mathrm{mm}}
\safemath{\cm}{\,\mathrm{cm}}
\safemath{\m}{\,\mathrm{m}}
\safemath{\W}{\,\mathrm{W}}
\safemath{\mW}{\, \mathrm{mW}}
\safemath{\J}{\,\mathrm{J}}
\safemath{\K}{\,\mathrm{K}}
\safemath{\bit}{\,\mathrm{bit}}
\safemath{\nat}{\,\mathrm{nat}}

\safemath{\define}{\triangleq}			

\safemath{\equivalent}{\sim}
\safemath{\distas}{\sim}					
\safemath{\sdiff}{\Delta}				

\safemath{\reals}{\mathbb{R}}
\safemath{\positivereals}{\reals_{+}}
\safemath{\integers}{\mathbb{Z}}
\safemath{\posint}{\integers_{+}}
\safemath{\naturals}{\mathbb{N}}
\safemath{\posnaturals}{\naturals_{+}}
\safemath{\complexset}{\mathbb{C}}
\safemath{\rationals}{\mathbb{Q}}

\newcommand*{\fancyrefapplabelprefix}{app}		
\newcommand*{\fancyrefthmlabelprefix}{thm}		
\newcommand*{\fancyreflemlabelprefix}{lem}		
\newcommand*{\fancyrefcorlabelprefix}{cor}		
\newcommand*{\fancyrefdeflabelprefix}{def}		
\newcommand*{\fancyrefalglabelprefix}{alg}		
\newcommand*{\fancyrefproplabelprefix}{prop}		
\newcommand*{\fancyrefexmpllabelprefix}{exmpl}
\newcommand*{\fancyreftbllabelprefix}{tbl}
\frefformat{vario}{\fancyrefseclabelprefix}{Sec.~#1}
\frefformat{vario}{\fancyrefthmlabelprefix}{Thm.~#1}
\frefformat{vario}{\fancyreflemlabelprefix}{Lem.~#1}
\frefformat{vario}{\fancyrefcorlabelprefix}{Cor.~#1}
\frefformat{vario}{\fancyrefdeflabelprefix}{Def.~#1}
\frefformat{vario}{\fancyrefalglabelprefix}{Alg.~#1}
\frefformat{vario}{\fancyreffiglabelprefix}{Fig.~#1}
\frefformat{vario}{\fancyrefapplabelprefix}{App.~#1} 
\frefformat{vario}{\fancyrefeqlabelprefix}{(#1)}
\frefformat{vario}{\fancyrefproplabelprefix}{Prop.~#1}
\frefformat{vario}{\fancyrefexmpllabelprefix}{Ex.~#1}
\frefformat{vario}{\fancyreftbllabelprefix}{Tbl.~#1}

\safemath{\dictab}{[\,\dicta\,\,\dictb\,]}

\safemath{\ysig}{\bmy}
\safemath{\ysighat}{\hat{\ysig}}
\safemath{\ysigdim}{M}
\safemath{\xsig}{\bmx}
\safemath{\xsigdim}{N}
\safemath{\nx}{n_x}
\safemath{\zsig}{\bmz}
\safemath{\zsigdim}{\ysigdim}
\safemath{\rsig}{\bmr}
\safemath{\Adict}{\bA}
\safemath{\Adicttilde}{\widetilde{\Adict}}
\safemath{\Adictdim}{\outputdim\times\xsigdim}
\safemath{\avec}{\bma}
\safemath{\avectilde}{\tilde{\avec}}
\safemath{\Bdict}{\bB}
\safemath{\Bdicttilde}{\widetilde{\Bdict}}
\safemath{\Cdict}{\bC}
\safemath{\cvec}{\bmc}
\safemath{\Ddict}{\bD}
\safemath{\Ddictdim}{\ysigdim\times\xsigdim}
\safemath{\dvec}{\bmd}
\safemath{\Ddicttilde}{\widetilde{\bD}}
\safemath{\Bonb}{\bB}
\safemath{\bvec}{\bmb}
\safemath{\Bonbdim}{\ysigdim\times\ysigdim}
\safemath{\noise}{\bmn}
\safemath{\noisedim}{\ysigim}
\safemath{\err}{\bme}
\safemath{\errdim}{\ysigdim}
\safemath{\errset}{\setE}
\safemath{\nerr}{n_e}
\safemath{\delop}{\bP_\errset}
\safemath{\delopc}{\bP_{{\errset}^c}}

\safemath{\cplxi}{\imath}
\safemath{\cplxj}{\jmath}

\safemath{\dict}{\matD}
\safemath{\inputdim}{N}		
\safemath{\outputdim}{M}		
\safemath{\sparsity}{S}	
\safemath{\inputdimA}{{N_a}}	
\safemath{\inputdimB}{{N_b}}	
\safemath{\elemA}{{n_a}}	
\safemath{\elemB}{{n_b}}	
\safemath{\resA}{\matR_a}	
\safemath{\resB}{\matR_b}	
\safemath{\subD}{\matS} 
\safemath{\subA}{\matS_a} 
\safemath{\subB}{\matS_b} 
\safemath{\dicta}{\matA} 	
\safemath{\dictb}{\matB} 	
\safemath{\hollowS}{H}
\safemath{\hollowA}{H_a}
\safemath{\hollowB}{H_b}
\safemath{\cross}{Z}
\safemath{\coh}{\mu_d}			
\safemath{\coha}{\mu_a}			
\safemath{\cohb}{\mu_b}			
\safemath{\mubs}{\nu}	
\safemath{\cohm}{\mu_m} 
\safemath{\dictset}{\setD}	
\safemath{\dictsetp}{\dictset(\coh,\coha,\cohb)}	
\safemath{\dictsetgen}{\dictset_\text{gen}}
\safemath{\dictsetgenp}{\dictsetgen(\coh)}
\safemath{\dictsetonb}{\dictset_\text{onb}}
\safemath{\dictsetonbp}{\dictsetonb(\coh)}

\safemath{\leftside}{U}
\safemath{\rightsideA}{R_a}
\safemath{\rightsideB}{R_b}

\safemath{\indexS}{\setI_S} 

\safemath{\na}{n_a}			
\safemath{\nb}{n_b}			
\safemath{\coeffa}{p_i}	
\safemath{\coeffb}{q_j}	
\safemath{\seta}{\setP}		
\safemath{\setb}{\setQ}     
\safemath{\setw}{\setW}	
\safemath{\setz}{\setZ}	
\safemath{\cola}{\veca}		
\safemath{\colb}{\vecb}		
\safemath{\cold}{\vecd}		
\safemath{\inputvec}{\vecx} 	
\safemath{\error}{\vece}	
\safemath{\noiseout}{\vecz} 	
\safemath{\inputvecel}{x}
\safemath{\inputveca}{\vecx_a}
\safemath{\inputvecb}{\vecx_b}
\safemath{\outputvec}{\vecy}	
\safemath{\lambdamin}{\lambda_{\mathrm{min}}}

\safemath{\elltwo}{\ell_2}
\safemath{\ellone}{\ell_1}
\safemath{\ellzero}{\ell_0}
\safemath{\ellinf}{\ell_\infty}
\safemath{\licard}{Z(\coh,\coha,\cohb)}
\safemath{\xsol}{\hat{x}}
\safemath{\xbord}{x_b}		
\safemath{\xstat}{x_s}		
\safemath{\xstatLone}{\tilde{x}_s}
\safemath{\order}{\mathcal{O}} 
\safemath{\scales}{\Theta} 
\safemath{\ones}{\mathbf{1}} 
\safemath{\zeroes}{\mathbf{0}} 
\safemath{\thlone}{\kappa(\coh,\cohb)} 
\safemath{\constoneA}{\delta} 
\safemath{\constoneB}{\epsilon} 
\safemath{\nlarge}{L}				   
\safemath{\sumlarge}{S_\nlarge}
\safemath{\maxlarger}{P_\nlarge}	   
\safemath{\Pzero}{\textrm{P0}}	
\safemath{\Pone}{\textrm{P1}}
\safemath{\vecfir}{\vecw}			 
\safemath{\vecsec}{\vecz}
\safemath{\elvecfir}{w}              
\safemath{\elvecsec}{z}				 
\safemath{\nlargefir}{n}
\safemath{\normout}{\gamma}
\safemath{\auxfun}{h}
\safemath{\supp}{\textrm{supp}}

\safemath{\indexa}{\ell}
\safemath{\indexb}{r}
\safemath{\indexc}{i}
\safemath{\indexd}{j}

\safemath{\project}{P}

\ifCLASSINFOpdf

\else

\fi

\begin{document}
%
\title{{Insense: Incoherent Sensor Selection \\ for Sparse Signals}}


\author{\IEEEauthorblockN{Amirali Aghazadeh,\IEEEauthorrefmark{1}~\IEEEmembership{Student Member,~IEEE},
Mohammad Golbabaee,\IEEEauthorrefmark{2}
Andrew S.\ Lan,\IEEEauthorrefmark{3}  and \\
Richard G.\ Baraniuk,\IEEEauthorrefmark{1}~\IEEEmembership{Fellow,~IEEE}} \\
\IEEEauthorblockA{\IEEEauthorrefmark{1}Electrical and Computer Engineering Department, Rice University, Houston, TX 77005, USA} \\
\IEEEauthorblockA{\IEEEauthorrefmark{2}Institute for Digital Communication, School of Engineering, University of Edinburgh, Edinburgh EH9 3JL, UK}\\
\IEEEauthorblockA{\IEEEauthorrefmark{3}Princeton University, Electrical Engineering Department, Princeton, NJ, 08540, USA}
\thanks{Manuscript received XX XX, XX; revised XX XX, XX. 
Corresponding author: R.\ G.\ Baraniuk (email: richb@rice.edu).}}

\IEEEtitleabstractindextext{%
\begin{abstract}
Sensor selection refers to the problem of 
intelligently selecting a small subset of a collection of available sensors to reduce the sensing cost while preserving signal acquisition performance. 
The majority of sensor selection algorithms find the subset of sensors that best recovers an arbitrary signal from a number of linear measurements that is larger than the dimension of the signal. 
In this paper, we develop a new sensor selection algorithm for \emph{sparse} (or near sparse) signals that finds a subset of sensors that best recovers such signals from a number of measurements that is much smaller than the dimension of the signal.
Existing sensor selection algorithms cannot be applied in such situations.
Our proposed {\em Incoherent Sensor Selection (Insense)} algorithm minimizes a coherence-based cost function that is adapted from recent results in sparse recovery theory.
Using six datasets, including two real-world datasets on microbial diagnostics and structural health monitoring, we demonstrate the superior performance of Insense for sparse-signal sensor selection.

\end{abstract}

}

\maketitle

\IEEEdisplaynontitleabstractindextext
\IEEEpeerreviewmaketitle

\section{Introduction}
\label{sec:Intro}
%
%
%
%

\vspace{0cm}

The accelerating demand for capturing signals at high resolution is driving acquisition systems to employ an increasingly large number of sensing units.
However, factors like manufacturing costs, physical limitations, and energy constraints typically define a budget on the total number of sensors that can be implemented in a given system.
This budget constraint motivates the design of \emph{sensor selection} algorithms~\cite{joshi2009sensor} that intelligently select a subset of sensors from a pool of available sensors in order to lower the sensing cost with only a small deterioration in acquisition performance.

In this paper, we extend the classical sensor selection setup, where $D$ available sensors obtain linear measurements of a signal $x \in \mathbb{R}^N$ according to
$y = \Phi x$ with each row of $\Phi\in \mathbb{R}^{D\times N}$ corresponding to one sensor.
In this setup, the sensor selection problem is one of finding a subset $\Omega$ of sensors (i.e., rows of $\Phi$) of size $| \Omega | = M$ such that the signal $x$ can be recovered from its $M$ linear measurements 
\begin{align}\label{eq:cs_linear}
y_{\Omega} = \Phi_{\Omega}x
\end{align}
with minimal reconstruction error. 
Here, $\Phi_{\Omega}\in \mathbb{R}^{M \times N}$ is called the \emph{sensing matrix}; it contains the rows of $\Phi$ indexed by $\Omega$.

The lion's share of current sensor selection algorithms \cite{joshi2009sensor,shamaiah2010greedy,ranieri2014near} select sensors that best recover an arbitrary signal $x$ from $M>N$ measurements.  
In this case, (\ref{eq:cs_linear}) is \emph{overdetermined}. 
Given a subset of sensors $\Omega$, the signal $x$ is recovered simply by inverting the sensing matrix while computing $ \Phi_{\Omega}^{\dagger} y_{\Omega}$, where $\Phi_{\Omega}^{\dagger}$ is the pseudoinverse of $\Phi_{\Omega}$.

Such approaches do not exploit the fact that many real-world signals are (near) \emph{sparse} in some basis \cite{candes2008introduction}.
It is now well-known that (near) sparse signals can be accurately recovered from a number of linear measurements $M\ll N$ using sparse recovery/compressive sensing (CS) techniques
\cite{donoho2006compressed,baraniuk2007compressive,candes2006compressive}.
Conventional sensor selection algorithms are not designed to exploit low-dimensional signal structure.
Indeed, they typically fail to select the appropriate sensors for sparse signals in this \emph{underdetermined} setting ($M<N$).

\begin{figure}[t]
\vspace{- 0.0 cm}
\centering
\includegraphics[width=0.25\textwidth]{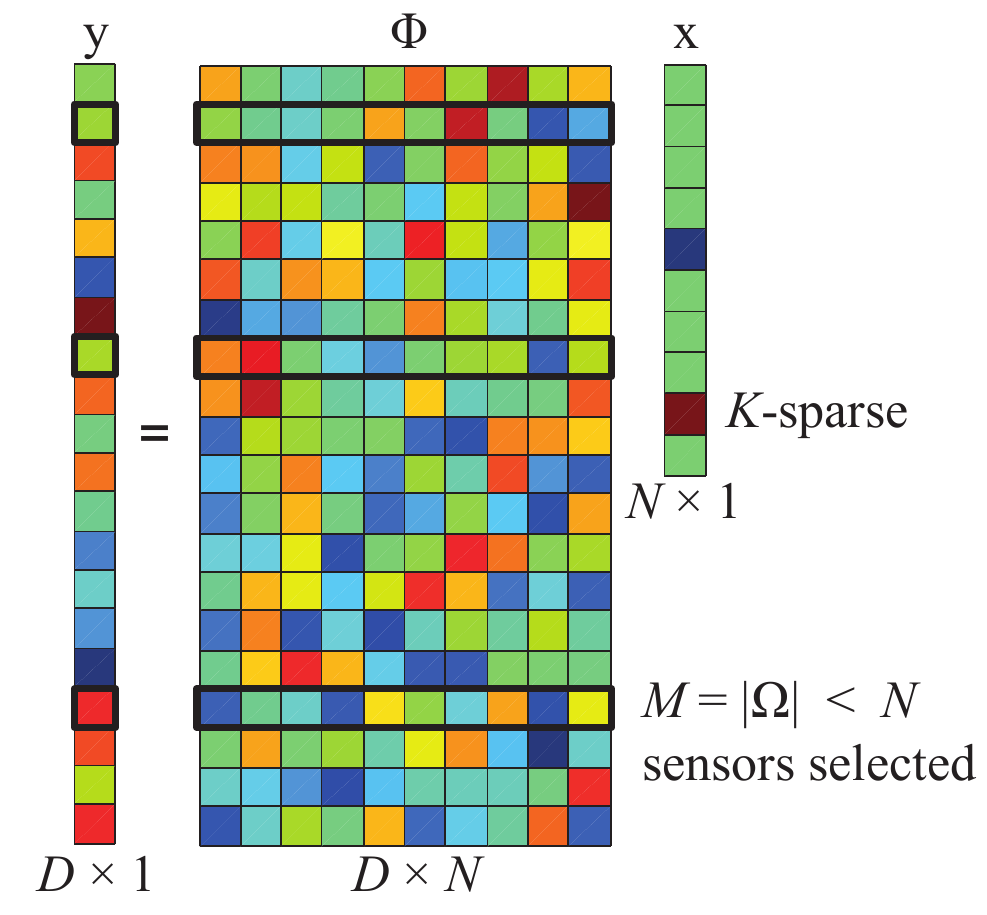}
\caption{Schematic of the sensor selection problem for sparse signals. Here, $M=3$ sensors indexed by $\Omega=\{2,8,17 \}$ are selected from $D=20$ available sensors to recover a $K=2$-sparse vector $x\in \mathbb{R}^{N}$, $N=10$, from the linear system $y_{\Omega} = \Phi_{\Omega}x$.  } 
\label{fig:spsene}
\vspace{- 0.5 cm}
\end{figure}

In this paper, we develop a new sensor selection framework that finds the optimal subset of sensors $\Omega$ that best recovers a (near) sparse signal $x$ from $M<N$ linear measurements (see Fig.~\ref{fig:spsene}). 
In contrast to the conventional sensor selection setting, here the sensing equation~\eqref{eq:cs_linear} is underdetermined, and it can not be simply inverted in closed form.

A key challenge in sensor selection in the underdetermined setting is that we must replace the cost function that has been so useful in the classical, overdetermined setting, namely the estimation error $\|x-\hat{x}\|_2^2$ (or the covariance of the estimation error in the presence of noise).
In the overdetermined setting, this error can be obtained in closed form simply by inverting~(\ref{eq:cs_linear}).
In the underdetermined setting, this error has no closed form expression.
Indeed, recovery of a sparse vector $x$ from $y_{\Omega}$ requires a computational scheme  \cite{tibshirani1996regression,donoho2009message}.

Fortunately, the sparse recovery theory tells us that one can reliably recover a sufficiently sparse vector $x$ from its linear measurements $y_{\Omega}$ when the columns of the sensing matrix $\Phi_{\Omega}$ are sufficiently \emph{incoherent} \cite{tropp2004greed,donoho2006stable,herzet2013exact}. 
Define the \emph{coherence} between the columns $\phi_i$ and $\phi_j$ in the sensing matrix $\Phi_{\Omega}$ as
$
\mu_{ij}(\Phi_\Omega)  = \frac{| \langle \phi_i,\phi_j \rangle |}{\|\phi_i \| \|\phi_j \|}
$.   
If the values of $\mu_{ij}(\Phi_\Omega)$ for all pairs of columns $(i,j)$  are bounded by a certain threshold, then sparse recovery algorithms such as Basis Pursuit (BP) \cite{candes2006stable,gribonval2006exponential,tropp2004greed} can recover the sparse signal $x$ exactly.
This theory suggests a new cost function for sensor selection.
To select the sensors $\Omega$ that reliably recover a sparse vector, we can minimize the \emph{average squared coherence} 
\begin{align}\label{eq:muavg}
 \mu_{\text{avg}}^2(\Phi_\Omega) = \frac{1} {{N \choose 2}}  \underset{1\leq i < j \leq N}{\sum}{\mu_{ij}^2(\Phi_\Omega)}.
\end{align}
\noindent
The challenge now becomes formulating an optimization algorithm that selects the subset of the \emph{rows} of $\Phi$ (the sensors) whose {\em columns} have the smallest average squared coherence.

\subsection{Contributions} 

We make three distinct contributions in this work.
First, we are the first to propose and study the sparse-signal sensor selection problem. 
Below we demonstrate that the standard cost functions used in overdetermined sensor selection algorithms are not suitable for the underdetermined case.

Second, we develop a new sensor selection algorithm that optimizes the new cost function (\ref{eq:muavg}); call it the \emph{Incoherent Sensor Selection (Insense)} algorithm. 
Insense employs an efficient optimization technique to find a subset of sensors with smallest average coherence among the columns of the selected sensing matrix $\Phi_{\Omega}$.
The optimization technique -- projection onto the convex set defined by a scaled-boxed simplex (SBS) constraint  -- is of independent interest.
We have made the codes for the Insense algorithm available online at \href{https://github.com/amirmohan/Insense.git}{https://github.com/amirmohan/Insense.git}. 

Third, we demonstrate the superior performance of Insense over conventional sensor selection algorithms using an exhaustive set of experimental evaluations that include real-world datasets from microbial diagnostics and structural health monitoring and six performance metrics: average mutual coherence, maximum mutual coherence, sparse recovery performance, frame potential, condition number, and running time.
We demonstrate that, for the kinds of redundant, coherent, or structured $\Phi$ that are common in real-world applications, Insense finds the best subset of sensors in terms of sparse recovery performance by a wide margin.
Indeed, in these cases, many conventional sensor selection algorithms fail completely.

\subsection{Paper organization}
We first overview the state-of-the-art overdetermined sensor selection algorithms and elaborate on the unique properties of the undetermined problem and its relation to CS in Section \ref{sec:relatwork}. 
Section~\ref{sec:problemstatement} states our coherence minimization formulation to solve the underdetermined sensor selection problem, and Section~\ref{sec:incsense} details the Insense algorithm.
Section~\ref{sec:experiment} presents our simulation results on synthetic and real-world datasets, and Section~\ref{sec:conclusion} concludes the paper.

\section{Related work} \label{sec:relatwork}

This section overviews the conventional sensor selection algorithms and their cost functions for the overdetermined case ($M>N$).
We then detail the relationship of the underdetermined case to the field of CS.

\subsection{Conventional sensor selection algorithms}

Existing sensor selection algorithms mainly study the sensor selection problem in the overdetermined regime (when $M \geq N$) \cite{joshi2009sensor,shamaiah2010greedy,ranieri2012eigenmaps,ranieri2014near}.

In the overdetermined regime, robust signal recovery can be obtained using the solution to the least squares (LS) problem in the sensing model~\eqref{eq:cs_linear}, which motivates as a cost function the mean squared error (MSE) \cite{das2008algorithms,golovin2010online,das2011submodular} or a proxy of the MSE \cite{steinberg1984experimental,krause2008near,wang2004entropy} of the LS solution.

For instance, Joshi, et al.\ \cite{joshi2009sensor} employ a convex optimization-based algorithm to minimize the log-volume of the \emph{confidence ellipsoid} around the LS solution of $x$. 
Shamaiah et al.\ \cite{shamaiah2010greedy} develop a greedy algorithm that outperforms the convex approach in terms of MSE.
FrameSense \cite{ranieri2014near} minimize the frame potential (FP) of the selected matrix
\begin{align}\label{eq:fp}
 \text{FP}(\Phi_{\Omega}) = \sum_{ \forall (i,j)\, \in \, \Omega, \, i < j  } {| \langle \phi^i , \phi^j \rangle |^2},
\end{align} 
where $\phi^i$ represents the $i^\text{th}$ row of $\Phi$.
Several additional sensor selection algorithms that assume a non-linear observation model \cite{ford1989recent,chepuri2015sparsity} also operate only in the overdetermined regime.

\subsection{Connections to compressive sensing}

Our model for sensor selection has strong connections to, and enables powerful extensions of, the CS problem, in which a (near) sparse signal is recovered from a small number of randomized linear measurements
\cite{donoho2006compressed,baraniuk2007compressive,candes2006compressive}.
First, note that CS theory typically employs \emph{random} sensing matrices; for instance it has been shown that many ensembles of random matrices, including partial Fourier, Bernoulli, and Gaussian matrices, result in sensing matrices with guaranteed sparse recovery\cite{needell2009uniform,needell2010signal}.
Recently, there have been efforts to \emph{design} sensing matrices that outperform random matrices for certain recovery tasks \cite{amini2011deterministic,strohmer2003grassmannian,tropp2005designing,elad2007optimized,duarte2009learning}.
For instance, Grassmannian matrices \cite{strohmer2003grassmannian} attain the smallest possible mutual coherence and hence can lead to better performance in some applications.

However, many real-world applications do not involve random or Grassmannian sensing matrices; rather the sensing matrix is dictated by physical constraints that are specific and unique to each application. 
For example, in the sparse microbial diagnostic problem \cite{UMD}, the entries of the sensing matrix $\Phi$ are determined by the hybridization affinity of random DNA probes to microbial genomes and do not necessarily follow a random distribution.
Similarly, in the structural health monitoring problem \cite{balageas2006structural,zhou2015l1,SHM}, the sensing matrix is constrained by the solution to the wave equation.
A key outcome of this paper is a new approach to construct practical and realizable sensing matrices using underdetermined sensor selection (via Insense).


\section{Problem statement}\label{sec:problemstatement}

Consider a set of $D$ sensors taking nonadaptive, linear measurements of a $K$-sparse (i.e., with $K$ non-zero elements) vector $x \in \mathbb{R}^N$ following the linear system
$y = \Phi x$,
where $\Phi\in \mathbb{R}^{D\times N}$ is a given sensing matrix.
We aim to select a subset $\Omega$ of sensors of size $| \Omega | = M \ll D$, such that the sparse vector $x$ can be recovered from the resulting $M < N$ linear measurements 
$y_{\Omega} = \Phi_{\Omega}x$
with minimal reconstruction error. 
Here, $\Phi_{\Omega}$ contains the rows of $\Phi$ indexed by $\Omega$, and $y_{\Omega}$ contains the entries of $y$ indexed by $\Omega$. 
This model for the sensor selection problem can be adapted to more general scenarios.
For example, if the signal is sparse in a basis $\Psi$, then we simply consider $\Phi = \Theta \Psi$ as the new sensing matrix, where $\Theta$ is the original sensing matrix.

In order to find a subset $\Omega$ of sensors (rows of $\Phi$) that best recovers a sparse signal $x$ from $y_{\Omega}$,\footnote{Or find one of the solutions if many solutions exists.}
we aim to select a submatrix $\Phi_\Omega \in \mathbb{R}^{M\times N}$ that attains the lowest average squared coherence
\begin{align} \label{eq:muavg1}
\mu_{\text{avg}}^2(\Phi_\Omega) = \frac{1} {{N \choose 2}} \sum_{1\leq i < j \leq N}{\frac{| \langle \phi_i,\phi_j \rangle |^2}{\|\phi_i \|^2 \|\phi_j \|^2}},
\end{align} 
where $\phi_i$ denotes the $i^\text{th}$ column of $\Phi_\Omega$. 
The term $\mu_{\text{avg}}$ averages the off-diagonal entries of $\Phi_{\Omega}^T\Phi_{\Omega}$ (indexed by $1\leq i < j \leq N$) after column normalization.
Other measures of coherence (e.g., max coherence $\mu_{\text{max}}(\Phi_\Omega) = \underset{i<j}{\max} {\mu_{ij}}$) can also be employed by slightly modifying the optimization procedure developed below. 
We choose to work with average coherence due to its simplicity and the fact that our experiments show that its performance is comparable to max coherence.

Define the diagonal selector matrix $Z={\text{diag}}(z)$ with $z = [z_1,z_2,z_3,\dots, z_D]^T$ and $z_i\in\{0,1\}$, where $z_i=1$ indicates that the $i^{\text{th}}$ row (sensor) in $\Phi$ is selected and $z_i=0$ otherwise.
This enables us to formulate the sensor selection problem as the following optimization problem
\begin{align} 
\notag \underset{z \in \{0,1\}^D}{\text{minimize}} & \sum_{1\leq i < j \leq N} \frac{{G_{ij}}^2}{G_{ii} \,\,  G_{jj}}, \\
\text{s.t.} & \quad G = \Phi^{T} Z \Phi, \; \mathbf{1}^T z=M,
\label{eq:ssp1}
\end{align} 
where $\mathbf{1}$ is the all-ones vector. 
This Boolean optimization problem is combinatorial, since one needs to search over $D \choose M$ combinations of sensors to find the optimal set $\Omega$. 

To overcome this complexity, we relax the Boolean constraint on $z_i$ to the {\em box constraint} $z_i \in [0,1]$ to arrive at the following problem
\begin{align} 
\notag \underset{z\in [0,1]^D}{\text{minimize}} & \sum_{1\leq i < j \leq N}  \frac{{G_{ij}}^2}{G_{ii} \,\,  G_{jj}}, \\
\text{s.t.} & \quad G = \Phi^{T} Z \Phi, \; \mathbf{1}^T z=M,
\label{eq:problem1}
\end{align} 
which supports an efficient gradient--projection algorithm to find an approximate solution.
We develop this algorithm next.

\section{The Insense algorithm} \label{sec:incsense}
\label{sec:incsense}
We now outline the steps that Insense takes to solve the problem~\eqref{eq:problem1}.
We slightly modify the objective of \eqref{eq:problem1} to
\begin{align} \label{eq:smoothed}
f_{\epsilon}(z) = \sum_{1\leq i < j \leq N}  \frac{{G_{ij}}^2 + \epsilon_1}{G_{ii} \,\,  G_{jj}+\epsilon_2} \quad \text{where}\quad G = \Phi^{T} Z \Phi,
\end{align}
where the small positive constants $\epsilon_2 < \epsilon_1  \ll 1$ make the objective well-defined and bounded over $z\in [0,1]^D$.

The objective function \eqref{eq:smoothed} is smooth and differentiable but non-convex; the box constraints on $z$ are linear. 
We minimize the objective using the iterative gradient-projection algorithm outlined in Alg.~\ref{alg:insense}. 
The gradient $\nabla_{z} f \in \mathbb{R}^D$ can be computed using the multidimensional chain rule of derivatives \cite{petersen2008matrix} as
\begin{align*}
(\nabla_z f)_i = \left(\Phi \nabla_G f\Phi^T\right)_{ii}  \quad \text{for} \quad i=1,\ldots,D.
\end{align*}
The $N\times N$ upper triangular matrix $\nabla_G f$ is the gradient of $f$ in terms of the (auxillary) variable $G$ at the point $G = \Phi^T Z \Phi$, given by
\begin{align}\label{eq:gradient}
(\nabla_G f)_{ij} = \begin{dcases}
    \frac{2G_{ij}}{G_{ii} \,\,  G_{jj}+\epsilon_2} ,&  i < j\\[2mm]
   -\sum_{\forall \ell \neq i} \frac{G_{\ell \ell} \left(G^2_{i\ell} + \epsilon_1 \right) } { \left( G_{ii} \,\,  G_{\ell \ell}+\epsilon_2 \right)^2} ,              & i=j\\[2mm]
   0, &\text{elsewhere.}
\end{dcases} 
\end{align}

The Insense algorithm (Alg.~\ref{alg:insense}) proceeds as follows.
First, the variables $G$ and $Z$ are initialized.
Next, we perform the following update in iteration~$k$,
\begin{align}\label{eq:incsence seq}
z^{k+1} = P_{\text{SBS}}\left( z^k - \gamma^k \nabla_z f(z^k)\right),
\end{align}
where $P_{\text{SBS}}$ denotes the projection onto the convex set defined by the scaled boxed-simplex (SBS) constraints $\mathbf{1}^Tz=M$ and $z\in[0,1]^D$. 
For certain bounded step size rules (e.g., $\gamma^k \leq 1/L$, where $L$ is the Lipschitz constant of $\nabla_z f$), the sequence $\{ z^k \}$ generated by \fref{eq:incsence seq} converges to a critical point of the non-convex problem \cite{attouch2013convergence,nesterov2007gradient}. 
In our implementation, we use a backtracking line search to determine $\gamma^k$ in each step \cite{nesterov2007gradient}.

\subsection{The SBS projection}
\label{sec:proxm}
We now detail our approach to solving the SBS projection problem
\begin{align} \label{eq:prox}
\notag &\underset{z}{\text{minimize}} \quad \frac{1}{2}\| z-y \|_2^2, \\
&\text{subject to} \quad \sum_i z_i = M ~~ \text{and} ~~ z_i \in [0,1] ~~ \forall i = 1,\dots,D.
\end{align} 
We emphasize that, for $M>1$, the SBS projection problem is significantly different from the (scaled-)simplex constraint $\left(\sum_i z_i = M\right)$ projection problem that has been studied in the literature \cite{chen2011projection,wang2013projection,condat2014fast}, due to the additional box constraints $z_i \in [0,1]$.

The Lagrangian of the problem \eqref{eq:prox} can be written as
\begin{align*} 
&f(z,\lambda, {\alpha}, {\beta}) =  \frac{1}{2}\| z-y \|_2^2 \\
& \quad +\lambda \left( \sum_i z_i - M \right) + \sum_i \alpha_i (-z_i) + \sum_i \beta_i (z_i - 1), 
\end{align*} 
where $\lambda,{\alpha},{\beta}$ are Lagrange multipliers for the equality and inequality constraints, respectively. The Karush-Kuhn-Tucker (KKT) conditions are given by
\begin{align*} 
& z_i - y_i + \lambda - \alpha_i + \beta_i = 0, \forall i, \\
& \sum_i z_i - M = 0, \\
& \alpha_i (-z_i) = 0, \, \beta_i (z_i - 1) = 0, \, \alpha_i,\beta_i \geq 0, \, 0 \leq z_i \leq 1, \, \forall i.
\end{align*}
According to the complimentary slackness condition for the box constraint $z_i \in [0,1]$, we have the following three cases for $x_i$:
\begin{itemize}
\item[(a)] $z_i = 0$: $\beta_i = 0, \alpha_i = y_i + \lambda > 0$,
\item[(b)] $z_i = 1$: $\alpha_i = 0, \beta_i = 1- y_i - \lambda > 0$,
\item[(c)] $z_i \in (0,1)$: $\alpha_i = \beta_i = 0, z_i = y_i + \lambda$.
\end{itemize}

Therefore, the value of $\lambda$ holds the key to the proximal problem \fref{eq:prox}. 
However, finding $\lambda$ is not an easy task, since we do not know which entries of $z$ will fall on the boundary of the box constraint (and are equal to either $0$ or $1$).

\begin{algorithm}[t]  \label{alg:insense}
\:
\SetAlgoLined
{\bf Input}: {$\Phi$} \\
{\bf Output}: {$Z = \text{diag}(z)$} \\
{\bf Initialization}: \\
 ${z} \leftarrow {z_{0}}$; \\
 ${G} \leftarrow {\Phi^T Z \Phi}$;

\While{ \text{stopping criterion = false} }{
1. $k \leftarrow k+1$\;
2. update $\nabla_z f(z^k)$ based on equation (\ref{eq:gradient})\;
3. $\gamma_k \leftarrow \text{line search}(f,\nabla_z f(z^k) ,z^k)$\; 
4. $ z^k \leftarrow z^k - \gamma^k \nabla_z f(z^k)$ \{gradient step\}\; 
5. $z^{k+1} \leftarrow P_{\text{SBS}}(z^k)$ \{SBS projection step\}\; 
}
\caption{Insense}
\end{algorithm} 
\vspace{-0.0cm}

In order to find the entries $z_i$ that are equal to $0$, we assume without loss of generality that the values of $y$ are sorted in ascending order: $y_1 \leq y_2 \leq \ldots y_D$. 
We note that, in all three cases above, $z_i = \max(\min(y_i + \lambda,1),0)$. 
Therefore, $\sum_i z_i$ is a non-decreasing function of $\lambda$. 
We evaluate $\sum_i z_i$ at the following values of $\lambda$:
\begin{itemize}[leftmargin=*]
\item $\lambda = -y_1$: $z_1 = 0, z_i = \min(y_i - y_1,1)$ for $i \geq 2$,
\item $\lambda = -y_2$: $z_1 = z_2 = 0$, $z_i = \min(y_i - y_1,1)$ for $i \geq 3$, \\
\ldots
\item $\lambda = -y_D$: $z_1 = z_2 = \ldots = z_D = 0$.
\end{itemize}
Thus, the entries in $z$ that are equal to $0$ correspond to the first $K_0$ entries in $y$, where $K_0$ is the largest integer $k$ such that $\sum_i \max(\min(y_i - y_k,1),0) \geq M$. 

Similarly, we can find the entries in $z$ that are equal to $1$ by negating $z$ and $y$ in \fref{eq:prox}. 
Let $p = -y$ and assume that its entries are sorted in ascending order; a procedure similar to that above shows that the entries in $z$ that are equal to $1$ correspond to the first $K_1$ entries in $p$, where $K_1$ is the largest integer $K$ such that $\sum_i \max(\min(p_i - p_k -1,0),-1) \geq - M$. 

Knowing which entries in $z$ are equal to $0$ and $1$, we can solve for the value of $\lambda$ by working with the entries with values in $(0,1)$. Using case (c) above and denoting the index set of these entries by $\zeta$, we have 
\begin{align*} 
\lambda = \frac{M - K_1 - \sum_{i \in \zeta} y_i}{|\zeta|},
\end{align*}
and the solution to \fref{eq:prox} is given by $z_i = \max(\min(y_i + \lambda,1),0)$.

\section{Experiments} \label{sec:experiment}
\label{sec:expe}

In this section, we experimentally validate Insense (Alg.~\ref{alg:insense}) using a range of synthetic and real-world datasets. 
In all experiments, we set $\epsilon_1 = 10^{-9}$ and $\epsilon_2 = 10^{-10}$
(anything in the range $\epsilon_2 < \epsilon_1  \ll 1$ can be utilized).
We terminate Insense when the relative change of the cost function $\mu_{\text{avg}}^2(\Phi_{\Omega})$ drops below $10^{-7}$.

\subsection{Baselines and performance metrics}

We compare Insense with several leading sensor selection algorithms, including \emph{Convex Sensor Selection} \cite{joshi2009sensor}, \emph{Greedy Sensor Selection} \cite{shamaiah2010greedy}, \emph{EigenMaps} \cite{ranieri2012eigenmaps}, and \emph{FrameSense} \cite{ranieri2014near}.
We also compare with four greedy sensor selection algorithms that were featured in \cite{ranieri2014near}.
The first three minimize different information theoretic measures of the selected sensing matrix as a proxy to the MSE:  the determinant in \emph{Determinant-G} \cite{steinberg1984experimental}, mutual information (MI) in \emph{MI-G} \cite{krause2008near}, and entropy in \emph{Entropy-G} \cite{wang2004entropy}.
The final greedy algorithm, \emph{MSE-G} \cite{das2008algorithms,golovin2010online,das2011submodular}, directly minimizes the MSE of the LS reconstruction error.
The code for these baseline greedy algorithms were obtained from \href{https://github.com/jranieri/OptimalSensorPlacement}{https://github.com/jranieri/OptimalSensorPlacement}. 
We also compare with \emph{Random}, a simple baseline that selects sensors at random.

\begin{figure*}[t]
\vspace{-0.4cm}
\centering
\includegraphics[width=0.6\textwidth]{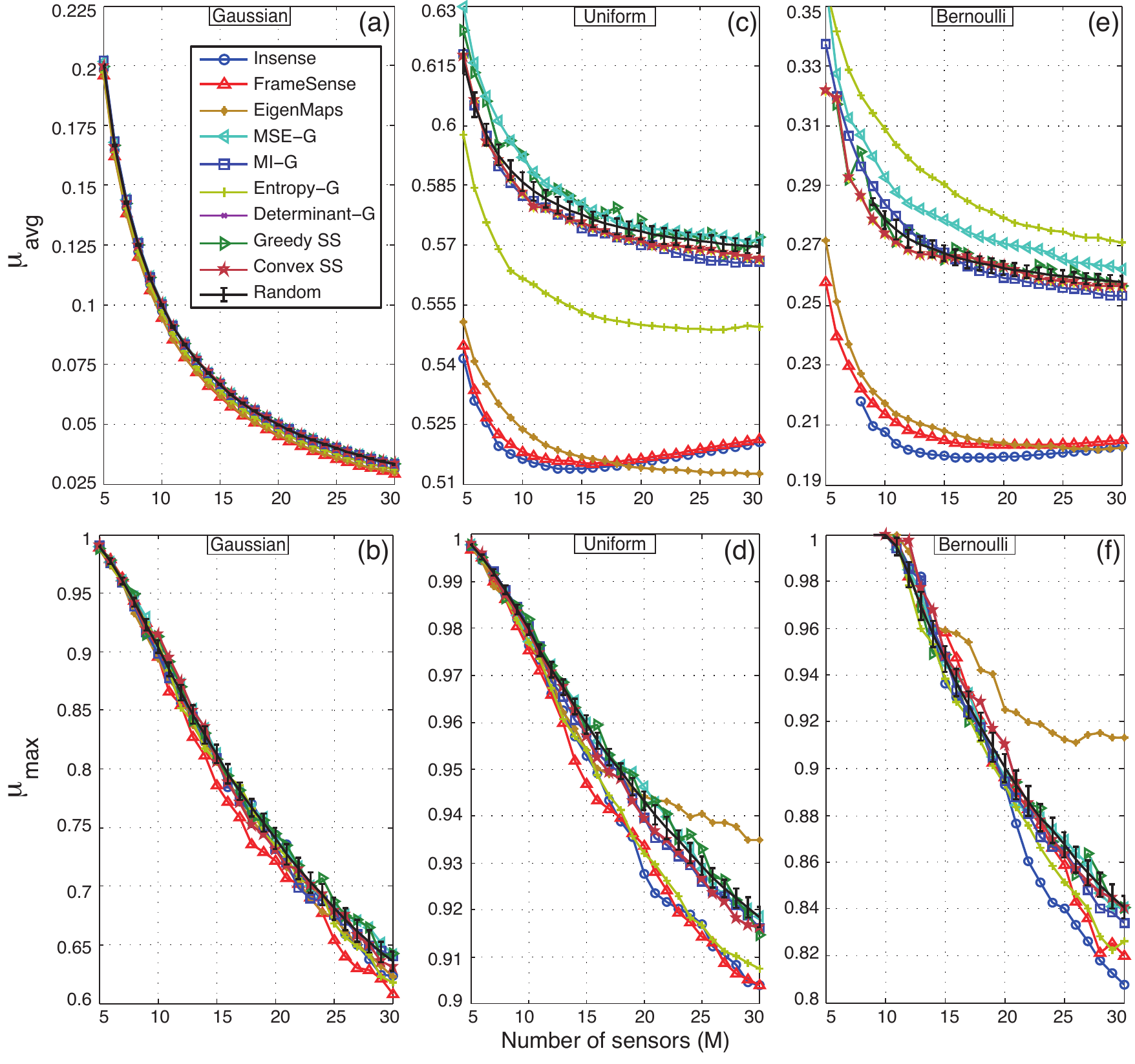}
\vspace{- 0.3 cm}
\caption{Comparison of Insense against the baseline algorithms in minimizing the average coherence $\mu_{\text{avg}}$ and maximum coherence $\mu_{\text{max}}$ of the selected sensing matrix $\Phi_{\Omega}$ from random sensing matrices with independent Gaussian (a,b), Uniform (c,d), and Bernoulli (e,f) entries ($D = N = 100$). 
Results are averaged over 20 trials with different random matrices.} 
\label{fig:GBU}
\vspace{-0cm}
\end{figure*}

We compare the sensor selection algorithms using the following six metrics:
\begin{itemize}[leftmargin=*]
\item \emph{Average coherence} $\mu_{\text{avg}}(\Phi_\Omega)$.
\item \emph{Maximum coherence} $\mu_{\text{max}}(\Phi_\Omega)$.
\item \emph{Frame potential} FP($\Phi_\Omega$) (see (\ref{eq:fp}))
\item \emph{Condition number} CN($\Phi_\Omega$).
\item \emph{BP recovery accuracy}. 
\item \emph{Running time} 
\end{itemize}
Depending on the task, in some experiments we only report a subset of the metrics.
To compute BP recovery accuracy, we average the performance of the BP algorithm \cite{chen1998atomic} over multiple trials.
In each trial we first generate a $K$-sparse vector $x$ whose non-zero entries are equal to one. We then use BP to recover $x$ from linear, nonadaptive (noiseless) measurements $y = \Phi_{\Omega} x$.

We repeat the same experiment for all $N \choose K$ sparse vectors $x$ with different support sets and report the percentage of trials that $x$ has been exactly recovered.
When $N \choose K$ is too large (here, greater than 10,000), we run the BP algorithm on a smaller random subset of all $N \choose K$ sparse vectors $x$.

\subsection{Unstructured synthetic datasets}

We first test the sensor section algorithms by applying them to random matrices.
It is easy to show that asymptotically (i.e., when $N \rightarrow \infty$) random matrices do not favor certain rows (sensors) over others.
In the non-asymptotic regime (i.e., when $N$ is finite) the choice of sensors for sparse recovery might be critical, since the probability that certain sets of sensors significantly outperform others increases. 
In this case, the sensor selection algorithm aims to identify these high-performing sensors.
We generate three types of random sensing matrices $\Phi$ whose entries are drawn from Gaussian, Uniform, and Bernoulli distributions and compare the performance of Insense against the other baselines.

\subsubsection{Random Gaussian matrix}

We conduct this experiment for 20 random trials and generate $100 \times 100$ matrices $\Phi$ whose entries are independently drawn from a standard normal distribution.
We use Insense and other baseline algorithms to select $M \in \{5,6,7,\ldots, 30\}$ sensors.
In Fig.~\ref{fig:GBU}(a,b) we report $\mu_{\text{avg}}$ and $\mu_{\max}$ of the selected sub-matrices 
$\Phi_{\Omega}$, with $|\Omega|=M$.
All of the sensor selection algorithms have comparable performance to the random sensor selection strategy (Random), illustrating that only small improvements to the maximum and average coherence can be achieved using these algorithms.

\subsubsection{Random Uniform matrix}

We repeat the previous experiment with a sensing matrix $\Phi$ whose entries are drawn uniformly at random from $[0,1]$. 
Fig.~\ref{fig:GBU}(c)) shows that Insense outperforms most of the baseline algorithms, including Random and Convex SS, in terms of $\mu_{\text{avg}}$.
Despite selecting completely different sensors, FrameSense and EigenMaps have comparable performance to Insense in minimizing $\mu_{\text{avg}}$.
Fig.~\ref{fig:GBU}(d) makes apparent the gap in maximum coherence $\mu_{\text{max}}$ between that achieved by Insense and the other baselines.

\subsubsection{Random Bernoulli matrix}

We repeat the previous experiment with a sensing matrix $\Phi$ whose entries are 0 or 1 with equal probability.\footnote{In our experiments, the coherence minimization performance of all of the sensor selection algorithms was similar for Bernoulli ($1$/-$1$) matrices and Gaussian matrices.}
Fig.~\ref{fig:GBU}(e) shows that FrameSense, EigenMaps, and Insense have similar performance and outperform the other algorithms by a large margin in terms of average coherence $\mu_{\text{avg}}$.\footnote{When a selected matrices $\Phi_{\Omega}$ contains one column with all zero entries, the average coherence $\mu_{\text{avg}}$ is not defined. The missing values in some curves correspond to these instances.}
Fig.~\ref{fig:GBU}(f) shows a clear gap between Insense and the other baselines in terms of the maximum coherence $\mu_{\text{max}}$.

\begin{table*}[!]
\vspace{0cm}
\centering
\caption{Comparison of Insense against the baseline algorithms on selecting $M=10$ rows from a structured Identity/Gaussian $\Phi$.  Insense selects the set of sensors with the smallest $\mu_{\text{avg}}$ and achieves the best BP recovery performance.}
\scalebox{1}{%
\label{tab:randomiden}
\begin{tabular}{lcccc}
\toprule[0.2em]
 & $\mu_{\text{avg}(\Phi_{\Omega})}$ & FP($\Phi_{\Omega}$) & CN($\Phi_{\Omega}$) & BP accuracy \% \\
\midrule[0.1em]
Insense & {\bf 0.3061 $\pm$ 0.0047 } &  1019  $\pm  313$  & 1.93 $\pm$ 0.19 & {\bf 92.27 $\pm$ 1.42}\\
FrameSense & -- & {\bf0.00 $\pm$ 0.00}  & {\bf 1.00 $\pm$ 0.00} & 4.00 $\pm$ 0.00 \\ 
EigenMaps & -- & {\bf0.00 $\pm$ 0.00} & {\bf 1.00 $\pm$ 0.00} & 4.00 $\pm$ 0.00 \\ 
MSE-G & 0.3872 $\pm$ 0.0305 & 1155 $\pm$ 374 & 11.51 $\pm$ 0.93 & 57.91 $\pm$ 1.09\\
MI-G & -- & {\bf0.00 $\pm$ 0.00} & {\bf 1.00 $\pm$ 0.00}  & 4.00 $\pm$ 0.00\\ 
Entropy-G & -- & {\bf0.00 $\pm$ 0.00} & {\bf 1.00 $\pm$ 0.00} & 4.00 $\pm$ 0.00 \\ 
Determinant-G & -- & {\bf0.00 $\pm$ 0.00}  & {\bf 1.00 $\pm$ 0.00} & 4.00 $\pm$ 0.00 \\ 
Greedy SS & -- & {\bf0.00 $\pm$ 0.00} & {\bf 1.00 $\pm$ 0.00} & 4.00 $\pm$ 0.00 \\ 
Convex SS & 0.3137 $\pm$ 0.0075 & 2279 $\pm$ 470 & 2.22 $\pm$ 0.25 & 88.64 $\pm$ 3.64\\
\bottomrule[0.2em]\\
\end{tabular}
}
\vspace{-0.5cm}
\end{table*}

In summary, Insense selects reliable sensors that are consistently better than or comparable to the other baseline algorithms on random sensing matrices.
This suggests that Insense could find application in designing sensing matrices that outperform random matrices for CS recovery tasks.

\subsection{Highly structured synthetic datasets}

In contrast to random matrices, the sensing matrices in real-world applications often have imposed structures or redundancies.
In such cases, careful sensor selection can mean the difference between low and high performance.  
We now explore sensor selection with structured over-complete matrices by constructing two synthetic datasets that resembles the redundancies and structures in real-world datasets. 
Similar over-complete basis has been explored in \cite{chen1998atomic}.

\subsubsection{Identity/Gaussian matrix}

We construct our first highly structured dataset by concatenating two $50\times 50$ matrices:
An identity matrix and a random matrix with i.i.d. Gaussian entries. 
Such matrices feature prominently in certain real-world CS problems.

For instance, in the universal DNA-based microbial diagnostics platform studied in \cite{UMD}, the identity and Gaussian matrices symbolize two different types of sensors: the identity matrix corresponds to a set of sensors that are designed to be specific to a single microbial target (column) in the dictionary $\Phi$, while the Gaussian matrix corresponds to a set of sensors that are universal for microbial targets in the dictionary.
(We extend this simulation example with real experimental data below in Section \ref{sec:microb}.)
As in \cite{UMD}, consider a bacterial detection scenario where the solution to the sparse recovery problem both detects and identifies the bacterial targets in a sample (through the support of the sparse vector $x$); here the goal is to maximize the average sparse recovery (detection) performance.
On the one hand, if all of the sensors are selected from the identity submatrix, then nearly all of the selected sensors will  lie dormant when detecting a particular bacterial target.
On the other hand, if the sensors are selected from the Gaussian submatrix, then the selected sensors will work jointly to detect all bacterial targets, which provides both universality and better average sparse recovery performance \cite{UMD}.
To achieve a better sparse recovery performance, the sensor selection algorithm should select rows (sensors) from the Gaussian submatrix rather than the identity submatrix.

Table~\ref{tab:randomiden} compares the performance of Insense to the baseline algorithms for the problem of selecting $M=10$ rows from the structured Identity/Gaussian $\Phi$. 
We repeat the same experiment 10 times with different random Gaussian matrices.\footnote{Dashes correspond to instances where the selected matrices $\Phi_{\Omega}$ contain columns with all zero entries; here the average coherence $\mu_{\text{avg}}$ is undefined. }
In particular, Insense, Convex SS, and MSE-G are the only algorithms that select rows of the Gaussian sub-matrix.
While achieving the minimum FP($\Phi_{\Omega}$) ($=0$), the other algorithms perform poorly on BP recovery.
The greedy algorithms select rows from the identity matrix that result in columns with all-zero entries and thus fail to recover most of the entries in $x$. \
Digging deeper, Insense selects rows with smaller column coherence than Convex SS and MSE-G. 
As a result, Insense achieves the best BP recovery performance (Table~\ref{tab:randomiden}) among these three algorithms.

In summary, this example demonstrates that minimizing a similarity metric imposed on the rows of the sensing matrix (such as frame potential, etc.) will not maximize the recovery performance of sparse signals.
Our results also provide reassurance that the coherence among the columns of the sensing matrix is a useful performance objective.

\subsubsection{Uniform/Gaussian matrix}

To study the quality of the box constraint relaxation in (\ref{eq:problem1}), we compare Insense against the baseline algorithms for a matrix $\Phi$ where we know the globally optimal index set of rows (sensors) $\Omega$.
(For arbitrary $\Phi$, global combinatorial optimization is computationally intractable when $D,N>200$ or so.)

\begin{figure*}[t]
\centering
\scalebox{0.7}{
\includegraphics[width=2\columnwidth]{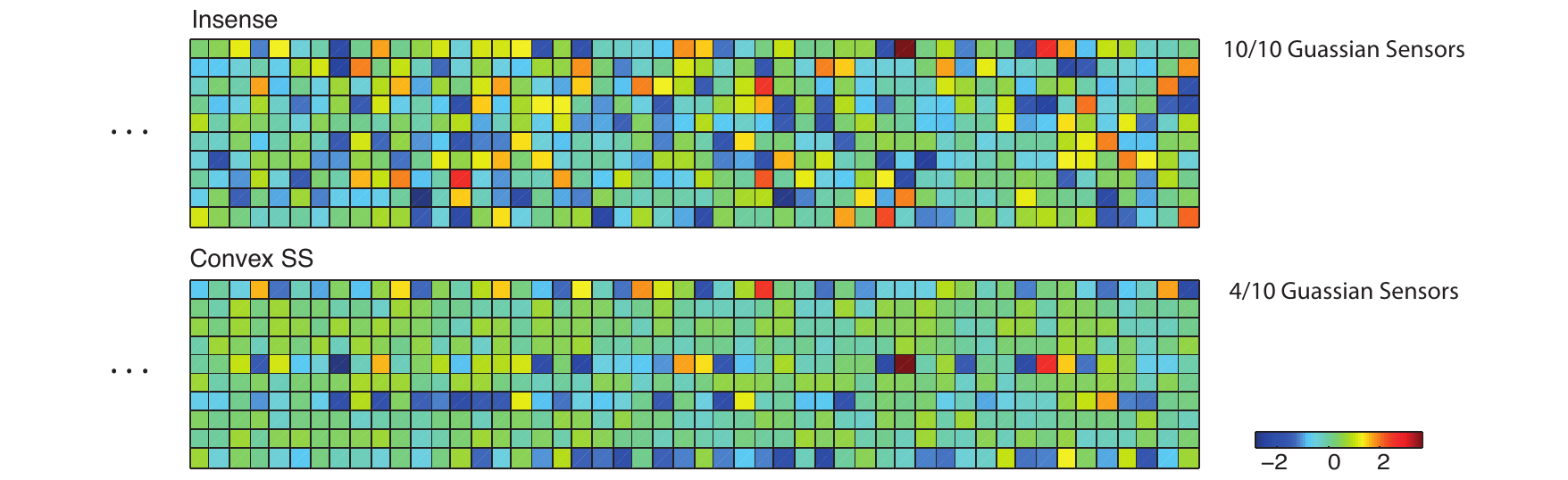}
}
\caption{Visualizations of the $M=10$ sensing matrices $\Phi_{\Omega}$ selected by Insense and Convex SS 
from a structured Uniform/Gaussian matrix.  
Insense selects 10/10 Gaussian rows (sensors), while Convex SS selects only 4/10 Gaussian rows. 
}
\label{fig:GU}
\vspace{-0.0cm}
\end{figure*}

We concatenate a $10\times 200$ matrix with i.i.d. Gaussian entries and a $190\times 200$ matrix with i.i.d. $[0,1]$ uniform distribution entries.
In this case, one would expect that the Gaussian submatrix has the lowest $\mu_{\text{avg}}$ when we set $M=10$. 
Fig.~\ref{fig:GU} visualizes the results of running Insense and the Convex SS baseline algorithm on such a $\Phi$. 
In all 10 random trials, Insense successfully selects all Gaussian rows and hence find the globally optimal set of sensors. 
FrameSense and EigenMaps miss, on average, 10--20\% of the Gaussian sensors.
The other baselines algorithms, including Convex SS, select only a small portion ($<$20\%) of the Gaussian rows (sensors).
Table~\ref{fig:tableGU} also indicates that Insense achieves better BP recovery performance, since it selects  exclusively Gaussian rows, resulting in the minimum average coherence $\mu_{\text{avg}}$ of the resulting sensing matrix. 

\begin{table*}[!]
\vspace{0cm}
\centering
\caption{Comparison of Insense against the baseline algorithms on selecting $M=10$ rows from a structured Uniform/Gaussian $\Phi$.  Insense selects the set of sensors with the smallest $\mu_{\text{avg}}$ and achieves the best BP recovery performance.}
\scalebox{1}{%
\begin{tabular}{lccccc}
\toprule[0.2em]
 & $\mu_{\text{avg}}(\Phi_{\Omega})$ & FP($\Phi_{\Omega}$) & CN($\Phi_{\Omega}$) & Gaussian sensor ratio \% & BP accuracy \% \\
\midrule[0.1em]
Insense & {\bf 0.3165 $\pm$ 0.0023 } &  9320  $\pm  3292 $  & {\bf 1.46 $\pm$ 0.07} & {\bf 100 $\pm$ 0}  & {\bf 58.55 $\pm$ 2.64}\\
FrameSense & 0.3273 $\pm$ 0.0059 & {\bf 6095 $\pm$ 1708} & 3.19 $\pm$ 0.92 & 84 $\pm$ 5 & 58.15 $\pm$ 2.26 \\ 
EigenMaps & 0.3215 $\pm$ 0.0021 & 7230 $\pm$ 2319 &  2.07 $\pm$ 0.12 & 90 $\pm$ 0 & 57.60 $\pm$ 3.72\\
MSE-G & 0.5805 $\pm$ 0.0440 & 78530 $\pm$ 12450 & 5.99 $\pm$ 0.31 & 17 $\pm$ 4 & 49.90 $\pm$ 3.54\\
MI-G  & 0.6814 $\pm$ 0.0556 & 93260 $\pm$ 109250 & 6.26 $\pm$ 0.77 & 7 $\pm$ 4 & 51.60 $\pm$ 5.21\\
Entropy-G & 0.7007 $\pm$ 0.0804 & 98950 $\pm$ 16216 & 6.61 $\pm$ 0.48 & 5 $\pm$ 7 & 53.70 $\pm$ 5.21\\
Determinant-G  & 0.7303 $\pm$ 0.0545 & 105700 $\pm$ 11228 & 6.57 $\pm$ 0.31 & 3 $\pm$ 4 & 55.50 $\pm$ 4.50\\
Greedy SS & 0.7303 $\pm$ 0.0545 & 105700 $\pm$ 11228 & 5.57 $\pm$ 0.31 & 3 $\pm$ 4 & 55.50 $\pm$ 4.50\\
Convex SS & 0.5788 $\pm$ 0.1140 & 75270 $\pm$ 27383 & 5.97 $\pm$ 0.77 & 20 $\pm$ 15 & 54.40 $\pm$ 4.20\\
\bottomrule[0.2em]\\
\end{tabular}}
\label{fig:tableGU}
\end{table*}

\subsection{Real-world datasets}

Finally, we assess the performance of Insense on two real-world datasets from microbial diagnostics and structural health monitoring. 

\subsubsection{Microbial diagnostics}
\label{sec:microb}

\begin{table*}[!]
\vspace{0cm}
\centering
\caption{Comparison of Insense against the baseline algorithms on selecting $M$ DNA probes to identify pathogenic samples containing $K$ bacterial organisms. Insense selects the sets of DNA probes that achieve the best pathogen identification performance.}
\scalebox{0.95}{%
\begin{tabular}{lccccccccccccccc}
\toprule[0.2em]
 & \multicolumn{14}{c}{ BP accuracy in detecting organisms \% }   \\
\midrule[0.1em]
Number of organisms  & \multicolumn{5}{c}{ $K = 2$ } &  \multicolumn{4}{c}{ $K = 3$ } &  \multicolumn{5}{c}{ $K = 5$ }  \\
\midrule[0.1em]
Number of probes ($M$)  & $5$ & $8$ & $12$ & $15$ & & $8$ & $12$ & $15$ & $20$ & & $12$ &  $15$ & $20$ & $25$  \\
\midrule[0.1em]
Insense & 25.52 & \textbf{68.33} & \textbf{94.78} & \textbf{99.65} &&  \textbf{26.46} & \textbf{71.74} & \textbf{93.95} & \textbf{99.53} & & \textbf{16.78} &  \textbf{51.95} & \textbf{92.71} & \textbf{99.10}\\
FrameSense  & 27.73 & 61.83 & 88.40 & 95.71 & & 22.70 & 62.32 & 82.29 & 98.36 & &  10.79 & 35.16 & 81.92 & 96.50\\
EigenMaps  & 14.97 & 49.65 & 84.69 & 94.66 & & 13.17 & 54.68 & 78.09 & 96.25 & & 6.69 & 27.47 & 72.13 & 95.30 \\
MSE-G & 27.26 & 60.79 & 91.53 & 97.91 & & 22.01 &  67.16 & 89.15 & 98.40 & & 14.69 &  43.26 & 83.52 & 97.40\\
MI-G &  26.22 & 59.98 & 89.68 & 96.40 & & 20.96 & 65.69 & 84.10 & 97.39 & & 13.49 & 37.96 & 79.72 & 96.00\\
Entropy-G & \textbf{27.96} & 61.25 & 91.53 & 98.61 & & 21.51 & 66.35 & 88.96 & 99.19 & & 14.19 & 42.86 &  89.61 & 97.50\\
Determinant-G & 14.85 & 46.75 & 82.13 & 94.55 & & 12.49 & 48.97 & 76.13 & 96.03 & & 6.29 & 24.48 & 72.73 & 92.81 \\
Greedy SS & 25.52  &  57.54 &  87.70 &  96.87 & & 19.72 & 59.65 & 84.64 & 97.34 & &  10.99 & 36.16 & 80.22 & 94.11\\
Convex SS & 15.89  &  53.36 &  87.94 &  98.94 & & 14.29 & 57.58 & 87.59 & 98.89 & & 7.69 & 38.46 & 83.52 & 98.40 \\
Random & 25.57    &  61.53   &  88.79  &  96.66 & & 22.37 & 62.29 & 86.15 & 97.72 & & 12.79 & 38.88 & 82.94 & 86.44\\

\bottomrule[0.2em]\\
\end{tabular}}
\label{tab:DNA_accu}
\end{table*}

Microbial diagnostics seek to detect and identify microbial organisms in a sample. 
Next-generation systems detect and classify organisms using DNA probes that bind (hybridize) to the target sequence and emit some kind of signal (e.g., fluorescence).
Designing DNA probes for microbial diagnostics is an important application of sensor selection in the underdetermined sensing regime. 
For example, in the universal microbial sensing (UMD) framework \cite{UMD}, DNA probes acquire linear measurements from a microbial sample (e.g., bacterial, viral, etc.) in the form of a fluorescence resonance energy transfer (FRET) signal that indicates the hybridization affinity of a particular DNA probe to the organisms present in the sample.
Given a matrix $\Phi$ that relates the hybridization affinities of DNA probes to microbial species, the objective is to recover a sparse vector $x$ comprising the concentrations of the organisms in the sample from as few linear measurements as possible.

We run Insense and the baseline sensor selection algorithms on a large sensing matrix comprising the hybridization affinity of $D=100$ random DNA probes to $N=42$ bacterial species (as described in \cite{UMD}).
For each algorithm, after selecting $M$ probes and constructing a sensing matrix $\Phi_{\Omega}$ with $|\Omega|=M$, we perform BP recovery for multiple sparse vectors $x$ with random support (corresponding to the presence of a random subset of bacterial organisms).
We repeat the same experiment for all $\binom{N}{K}$ sparse vectors $x$ with $K=\{2,3,5\}$ non-zero elements (i.e., bacteria present) and report the average BP recovery performance on identifying the composition of the samples in Table~\ref{tab:DNA_accu}.
(To report the BP recovery for $K=5$ organisms, we randomly generate 1000 realizations of the sparse vector $x$ with $5$ active elements and average the BP recovery performance on selected samples.)

The DNA probes selected by Insense outperform all of the baseline algorithms in identifying the bacterial organisms present.
Specifically, Insense requires a smaller number of DNA probes than the other algorithm to achieve almost perfect detection performance (BP accuracy $>99\%$), suggesting that Insense is the most cost-efficient algorithm to select DNA probes for this application.
Moreover, the performance gap between Insense and the other algorithms grows as the number of bacterial species present in the sample $K$ increases, indicating that Insense has better recovery performance in complex biological samples.

\subsubsection{Structural health monitoring}
\label{sec:shm}

\begin{figure*}[t]
\centering
\scalebox{0.8}{
\includegraphics[width=2\columnwidth]{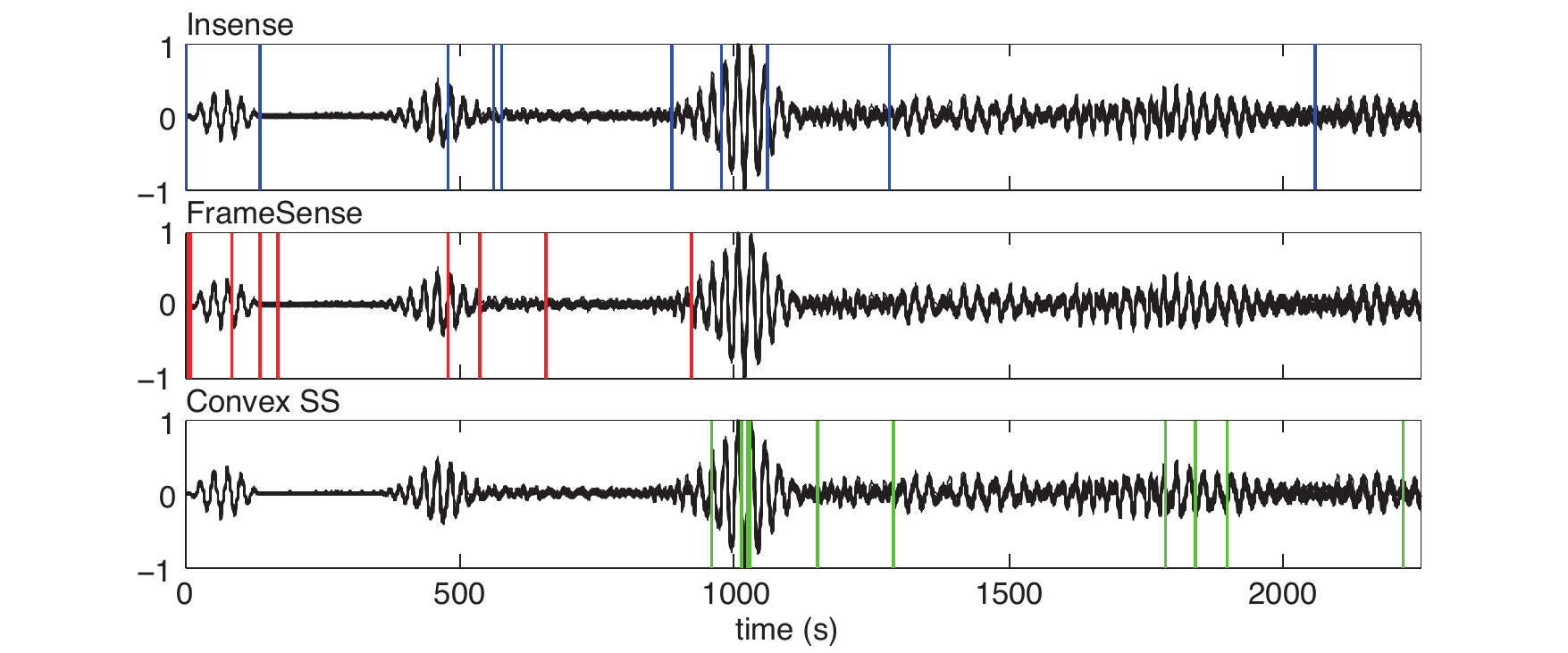}
}
\caption{Visualizations of the $M=10$ time instances selected by Insense (blue lines), FrameSense (red lines), and Convex SS (red lines) from a matrix for the crack localization problem.  
Insense selects time instances that are more distributed along the time axis and variant along the intensity axis. 
}
\label{fig:SHM_fig}
\vspace{-0.0cm}
\end{figure*}
\begin{table*}[!]
\vspace{0cm}
\centering
\caption{Comparison of Insense against the baseline algorithms on sampling $M$ time instances that best localize cracks on a plane in a structural health monitoring problem. Insense selects the set of sensors with the smallest $\mu_{\text{avg}}$ and achieves the best damage localization performance.}
\scalebox{0.9}{%
\begin{tabular}{lccccccccccccccc}
\toprule[0.2em]
  & \multicolumn{7}{c}{ $\mu_{\text{avg}}(\Phi_{\Omega})$ } &  \multicolumn{7}{c}{ BP accuracy in localizing cracks \% } &  \multicolumn{1}{c}{ $\text{Running time (s)}$ }  \\
\midrule[0.1em]
Number of samples ($M$)   & $5$ & $6$ & $7$ & $8$ & $9$ & $10$ & & $5$ & $6$ & $7$ & $8$ & $9$ & $10$ & & $10$ \\
\midrule[0.1em]

Insense & \textbf{0.2491} & \textbf{0.2251} & \textbf{0.2185} & \textbf{0.2127} &  \textbf{0.2075} & \textbf{0.2030} & & \textbf{40.99} & \textbf{60.49} & \textbf{76.49} &  \textbf{86.09} & \textbf{91.39} & \textbf{94.64} & & 4.12\\
FrameSense  & -- & 0.4051 & 0.3843 & 0.3179 & 0.2717 & 0.2640 & & 24.41 &  51.58 & 63.73 & 81.92 & 89.04 & 91.14 & & 0.18 \\
EigenMaps & 0.4341 & 0.4102 & 0.4028 & 0.3950 & 0.3841 & 0.3699 & & 25.13 & 41.16 &  40.55 & 51.38 & 64.52 & 80.45 & & 0.11 \\
MSE-G & 0.9832 & 0.9779 & 0.9752 & 0.9748 & 0.9707 &  0.9637 & & 39.50 & 53.50 & 63.73 &  79.18 & 86.28 & 93.50 & & 0.97\\
Entropy-G & 0.9694 & 0.9649 & 0.9615 & 0.9568 & 0.9525 & 0.9507 & & 40.89 & 58.95 & 71.35 & 80.98 &  86.50 & 92.36 & & 0.16\\
Greedy SS & --  &  -- &  -- &  -- & -- & -- & & 9.32 & 14.96 &  19.55 & 28.36 & 45.77 & 57.43 & & 1.65\\
Convex SS & 0.9702  &  0.9685 &  0.9611 &  0.9571 & 0.9519 & 0.9451 & & 38.92 & 59.56 & 72.42 & 82.83 & 89.98 & 94.56 & & 4.52\\
Random & 0.7980 &  0.7976   &  0.8083  &  0.8147 & 0.8111 & 0.8132 & & 33.84 & 52.20 & 67.51 & 79.60 & 88.19 & 93.75 & & < $10^{-4}$ \\

\bottomrule[0.2em]\\
\end{tabular}}
\label{tab:SHM_table}
\end{table*}

Structural health monitoring (SHM) studies the problem of detecting, localizing, and characterizing damage (e.g., cracks, holes, etc.) occurring on or inside structures such as buildings, bridges, pipes, etc.\ \cite{balageas2006structural}.
Typical SHM systems continuously monitor a target structure by transmitting signals that propagate throughout the structure and then analyzing the reflected signals measured using an array of sensors. 
Classical methods for damage localization calculate the time-of-flight (TOF) and estimate the location of a crack using triangulation, which is computationally intensive.

Dictionary-based SHM methods sidestep the computational complexity of TOF methods by constructing a matrix of signal profiles, where each column corresponds to the impulse response of the structure with a crack at a certain location on a predefined grid \cite{zhou2015l1,SHM}. 
The locations of a small number of cracks in the structure are then determined by solving a sparse recovery problem, where the support of the sparse signal vector $x$ in equation (\ref{eq:cs_linear})  corresponds to the damage locations \cite{SHM}.
In dictionary-based SHM methods, each row of the sensing matrix $\Phi$ corresponds to a measured time instance of a reflected signals (see Fig.~\ref{fig:SHM_fig}).
Applying a sensor selection algorithms to the SHM matrix will reduce the number of measurements required while preserving the damage localization performance.

To compare Insense against the other baseline algorithms, we use real SHM data from a crack localization experiment on a metal plate with 25 potential cracks located on a $5\times 5$ square grid.
The experimentally measured signals of size $D = 2250$ from all cracks are obtained and stored in a matrix $\Phi \in \mathbb{R}^{2250\times25}$. 
The crack locations are then identified by solving the sensor selection problem with the experimentally measured sensing matrix $\Phi$.
Fig.~\ref{fig:SHM_fig} visualizes the $M=10$ selected time instances for Insense and two other baseline algorithms, and Table~\ref{tab:SHM_table} showcases their performance on BP recovery in localizing $K=2$ cracks using $M \in \{ 5,6,7,8,9,10\}$ time instances. 
MI-G and Determinant-G does not run on our desktop PC with 12 GB of RAM due to high memory requirements.
Insense ran  faster than the optimization-based algorithm Convex SS and only slightly slower than the greedy algorithms.

In summary, in this SHM application, Insense consistently outperforms the baseline algorithms in terms of crack localization accuracy and reduces the number of time measurement instances required for accurate localization.
We note that the measurements are not statistically independent in this dataset, since they correspond to measurements of the same reflected signal at different time instances, highlighting the superior performance of Insense in applications where the independence assumption in the sensing model is slightly violated.


\section{Conclusions}\label{sec:conclusion}

In this paper, we developed the Incoherent Sensor Selection (Insense) algorithm for the underdetermined sensor selection problem that optimizes the average squared coherence of the columns of the selected sensors (rows) via a computationally efficient relaxation.
Our synthetic and real-world data results have both verified the utility of the average squared coherence metric and the performance of the Insense algorithm.
In particular, Insense provides superior performance than existing state-of-the-art sensor selection algorithms, especially in the real-world problems of microbial diagnostics and structural health monitoring.

There are a number of avenues for future work.
One is finding computational shortcuts for operating on large-scale sensing matrices.
Another is exploring different variants of the Insense algorithm that select sensors under generalized sparse models, e.g., group-sparse, tree-sparse and low-rank models.

\balance

\section*{Acknowledgment}
We thank S. Nagarajaiah for providing the Structural Health Monitoring data. A. Aghazadeh, A. S. Lan, and R. G. Baraniuk were supported by National Science Foundation (NSF) grants CCF-1527501 and CCF-1502875, Defense Advanced Research Projects Agency (DARPA) Revolutionary Enhancement of Visibility by Exploiting Active Light-fields grant HR0011-16-C-0028, and Office of Naval Research (ONR) grant N00014-15-1-2735. M. Golbabaee was supported by the Swiss NSF research grant PBELP2-146721.     

\bibliographystyle{IEEEtran}
\bibliography{Insense}

\end{document}